# Exploit Prediction Scoring System (EPSS)

Jay Jacobs
jay@cyentia.com
Cyentia

Sasha Romanosky
sromanos@rand.org
RAND Corporation

Ben Edwards
ben@cyentia.com
Cyentia

Michael Roytman
michael@kennasecurity.com
Kenna Security

Idris Adjerid
iadjerid@vt.edu
Virginia Tech

Despite the massive investments in information security technologies and research over the past decades, the information security industry is still immature. In particular, the prioritization of remediation efforts within vulnerability management programs predominantly relies on a mixture of subjective expert opinion, severity scores, and incomplete data. Compounding the need for prioritization is the increase in the number of vulnerabilities the average enterprise has to remediate. This paper produces the first open, data-driven framework for assessing vulnerability threat, that is, the probability that a vulnerability will be exploited in the wild within the first twelve months after public disclosure. This scoring system has been designed to be simple enough to be implemented by practitioners without specialized tools or software, yet provides accurate estimates of exploitation. Moreover, the implementation is flexible enough that it can be updated as more, and better, data becomes available. We call this system the Exploit Prediction Scoring System, EPSS.

Keywords: EPSS, vulnerability management, exploited vulnerability, CVSS, security risk management,

Acknowledgements: The authors would like to sincerely thank Kenna Security and Fortinet for contributing vulnerability data.

# Introduction

Despite the massive investments in information security technologies and research over the past decades, the industry is still immature.[1] In particular, the ability to assess the risk of a software vulnerability relies predominantly on incomplete data, and basic characteristics of the vulnerability, rather than on data-driven processes and empirical observations. The consequences of this are many. First, it prevents firms and network defenders from efficiently prioritizing software patches, wasting countless hours and resources remediating vulnerabilities that could be delayed, or conversely delaying remediation of critical vulnerabilities. In addition, policy makers and government agencies charged with overseeing critical infrastructure sectors are unable to defensibly marshall resources or warn citizens about the potential changes in adversary threats from a newly discovered vulnerability.

A common approach to prioritizing vulnerability remediation is based on characterizing the severity of a given vulnerability, often by using the internationally recognized CVSS, the Common Vulnerability Scoring System (ITU-1521).[2] CVSS computes the severity of a vulnerability as a function of its characteristics, and the confidentiality, integrity, and availability impact to an information system. The CVSS specification is clear that the Base score, the most commonly used component, is not meant to reflect the overall risk. Consequently, it does not measure threat or the probability that a vulnerability will be used to attack a network.[3]

Nevertheless, it has been mis-interpreted by some organizations as a faithful measure of cyber security risk, and become a *de facto* standard when prioritizing remediation efforts. For example, the payment card industry data security standard (PCI-DSS) requires that vulnerabilities greater than 4.0 must be remediated by organizations storing or processing credit cards (PCI, 2018), and in 2019 the Department of Homeland Security (DHS) released a binding operational directive requiring federal agencies to remediate high and critical vulnerabilities according to the CVSS standard.[4]

While the vulnerability severity component is addressed by CVSS, a critical gap in the information security field is a proper measure and forecast of threat, which is what this research seeks to address.

---

[1] Estimates suggest a worldwide cyber security market over $170 billion dollars by 2024. See
https://www.marketwatch.com/press-release/cyber-security-market-size-is-expected-to-surpass-us-170-billion-by-2024-2019-04-23
[2] See https://www.itu.int/rec/T-REC-X.1521-201104-I/en.
[3] The Temporal metric group includes a metric for the presence of an exploit, it does not account for whether the vulnerability is actively being exploited.
[4] See Binding Operational Directive (BOD) 19-02, https://cyber.dhs.gov/bod/19-02/.

Specifically, we improve on previous work by Jacobs et al (2019) in a number of important ways. First, we employ a modeling technique (logistic regression) that is more transparent, intuitive and easily implemented. In any information security process, multiple stakeholders will necessarily need to understand and act on the output - IT operations, security, business stakeholders. A model therefore needs to be interoperable and a score change must be explainable and attributed to any one feature. Second, as inputs to the model we only use data that are publicly available, easily discoverable, or are otherwise unrestricted. Finally, we formalize the regression coefficients into a scoring system that can be automated and implemented with simple and widely available tools (such as a spreadsheet). Overall, we develop the first open, data-driven threat scoring system for predicting the probability that a vulnerability will be exploited within the 12 months following public disclosure.

We believe this scoring system has the opportunity for making a fundamental contribution to the information security field not just as a way to help network defenders more efficiently allocate resources, but also for policy makers in communicating the actual threat of computer vulnerabilities.

The next section discusses related literature, followed by a description of the datasets used in this research. We then present our estimating model, results, and formalize our probabilistic vulnerability scoring system. We conclude with a discussion on limitations and conclusion.

## Related Literature

This paper draws on a research related to the incentives and tradeoffs of information security investments, markets (criminal and otherwise), and processes related to understanding and improving how firms and organizations protect against cyber security threats. It also extends previous academic and industry computer and security-related scoring systems.

Most directly, this paper extends research by Jacobs et al (2019) which developed a machine learning model for estimating the probability that a vulnerability will be exploited. It also contributes to a growing body of specialized research that use different machine learning techniques to predict when a vulnerability will be exploited. While some papers use proof of concept (published exploit) code as their outcome measure (Bozorgi et al, 2010; Edkrantz and Said, 2015; Bullough et al, 2017; Almukaynizi et al, 2017) others use real-world exploit data (Sabottke et al, 2015), as is done in this paper.

Providing useful assessments of the threat of a given vulnerability (i.e. threat scoring) has been notoriously difficult, either from academics or industry coalitions. For example, while CVSS has become an industry standard for assessing fundamental characteristics of vulnerabilities, the Base score (and accompanying metrics) captures fundamental properties of a vulnerability and the subsequent impact to an information system from a successful attack. The Temporal metrics of CVSS, on the other hand, are meant to capture time-varying properties of a vulnerability, such as the maturity of exploit code, and available patching. However, the Temporal metrics have enjoyed much lower adoption, likely due to the fact that they require organizations to update the information, as well as made judgements based on threat intelligence or other exploit data sources, something which has proven difficult to accomplish. In addition, and most relevant to this work, while the metrics seek to capture notions of vulnerability threat, they are not informed by any data-driven or empirical estimates. These two limitations (absence of an authoritative entity to update the metric values, and lack of data to inform the score), are key contributions of this work.

In addition, Microsoft created the Exploitability Index in 2008[5] in order to provide customers information about the exploitability of vulnerabilities targeting Microsoft software.[6] The Threat index provides a qualitative (ordinal) rating from 0 ("exploitation detected" – the highest score), to 3 ("exploitation unlikely" – the lowest score). In 2011, the Index was updated to distinguish between current and past application versions, and to capture the ability for the exploit to cause either a temporary or repeated denial of service of the application. Importantly, this effort was a closed, Microsoft-only service, and so neither provided any transparency into it's algorithm, nor included exploitability scores for other software products. By contrast, this effort is designed to be an open and transparent scoring system that considers all platforms and applications for which there exist publicly known vulnerabilities (i.e. CVEs).

## Data

The data used in this research relates to vulnerabilities published in a two-year window between June 1, 2016 and June 1, 2018 which includes 25,159 vulnerabilities. Published vulnerability information is collected from MITRE's Common Vulnerability Enumeration (CVE) and includes the CVE identifier, description and references to external websites[7] discussing or describing the vulnerability. Data was also collected from NIST's National Vulnerability Database (NVD) and includes the CVSS base severity score, and the Common Platform Enumeration (CPE) information, which provides information about the affected products, platforms and vendors.

---

[5] Estimated launch date based on information from https://blogs.technet.microsoft.com/ecostrat/2008/11/13/one-month-analysis-exploitability-index/. Last accessed July 5th, 2019. Further, there is some discussion that the original index was specifically meant (or at least communicated) to capture the likelihood of exploit *within 30 days from patch release*, however, this qualification is not explicitly mentioned on the website.

[6] See https://www.microsoft.com/en-us/msrc/exploitability-index. Last accessed July 5th, 2019.

[7] See https://cve.mitre.org/data/refs/index.html

We also derived and assigned descriptive tags about each vulnerability by retrieving the text from references present in each CVE. We extracted common multiword expressions from the raw text using Rapid Automatic Keyword Extraction (Rose et al, 2010) and manually culled and normalized a list of 191 tags[8] encoded as binary features for each vulnerability.[9] Our goal was to capture attributes and concepts practitioners use in the field to describe vulnerabilities (such as "code execution" and "privilege escalation"). The manual review was necessary to normalize disparate phrasings of the same general concept (e.g. "SQLi" and "SQL injection" or a "buffer overflow" vs "buffer overrun").

We collect information about whether proof-of-concept exploit code or weaponized exploits exists. Exploit code was extracted from Exploit DB. While weaponized exploits were found by looking at the modules in Rapid 7's Metasploit framework, D2 Security's Elliot Framework and the Canvas Exploitation Framework.

The critical outcome variable, information about whether the vulnerability was exploited in the wild comes from a number of sources. Specifically, we collect data from Proofpoint, Fortinet, AlienVault and GreyNoise[10]. In total, we acquired 921 observations of recorded exploitations used in the wild within the window of study. This represents an overall exploitation rate of 3.7%. Extra care was taken in collecting this variable to ensure it was limited to exploitations within the first twelve months after the CVE was published. This requirement and availability of data across all the data sources is what established the window of study to be between June 1st, 2016 and June 1st, 2018 (we discuss more on this below).

# Feature Selection

We observe from our dataset that some variables are incredibly sparse (including some that are completely separable), or are highly correlated with other variables, and so the inclusion of these features, which could lead to biased model parameters, and unnecessary complexity (Allison, 2008). Therefore, we applied a 3-stage set of criteria for including features into the model. In Stage 1, we filtered out all variables where complete separation occurs (Allison, 2008). Since some data are highly unbalanced, many variables were never associated with (completely separated from) our minority class (exploited vulnerability) and these were removed. In Stage 2, we removed any variables that weren't observed in at least 1% of the vulnerability data. Finally, in Stage 3, we reviewed the list of variables with domain experts and removed variables in order to ensure the model didn't overfit to particular quirks of the dataset, and potentially introducing unnecessary bias. For example, several common web application vulnerabilities (cross-site scripting, cross-site request forgery, directory traversal, input validation) will rarely have intrusion detection signatures written for each individual vulnerability, as they are easy to implement and are detected with a 'class' of signatures rather than for a specific vulnerability. This is a consequence of the way intrusion detection signatures are written (or not written), and is an area for future work for modelling exploitation. Table 1 shows the initial count of variables from our raw data ("Raw Count"), for each category of variable, and the number of variables that met the requirements for each stage of filtering (complete separation, proportion, and expert review).

---

[8] Only 166 of the 191 matched the vulnerabilities within our sample.

[9] An alternative approach is the familiar 'bag of words' however we found this method to be less effective.

[10] We note that these are closed sources for the outcome variable. Our goal is for practitioners to be able to implement the model and calculate scores from open inputs. At this time fitting the model requires access to closed exploitation data.

| Category of variable | Raw Count | Stage 1: Complete Separation | Stage 2: Proportion | Stage 3: Expert Review |
|---|---|---|---|---|
| Tags | 166 | 111 | 46 | 35 |
| Vendors | 3374 | 171 | 16 | 15 |
| References | 45 | 28 | 16 | 1 |
| Weaponized code | 1 | 1 | 1 | 1 |
| POC code | 1 | 1 | 1 | 1 |

Table 1: Count of Variables through each stage of filtering.

## Estimating the probability of exploited in the next 12 months

Vulnerabilities are continually discovered and published. As a result, our data contain vulnerabilities of both varying age and durations until exploitation. To make the estimated probability normalized across all vulnerabilities, we established a 12-month time window after publication as a CVE, and only included vulnerabilities that had been observed for the full time period. We selected the 12-month time window because this was the single common time-window encoded across the multiple data sources we collected on exploited vulnerabilities.

In addition, this time period is empirically justified by research by Kenna Security and the Cyentia Institute which analyzed remediation data from almost 300 companies and discovered that the majority of vulnerabilities are remediated within the first year (median of 100 days, 75% of vulnerabilities were remediated within 392 days of discovery) (Kenna 2019). This suggests that the majority of remediation decisions and actions are taken within this time period.

## Estimating Model and Variable Selection

As described, our interest is to estimate a binary outcome model that will predict the probability of a vulnerability being exploited within 12 months of being publicly disclosed. Additionally, we also wish to create a scoring system that is 1) simple to implement 2) interpretable 3) parsimonious and 4) performant. We therefore begin with a standard logistic regression as our estimating model.

Including all possible features (variables) into a logistic regression -- while possible -- would not be parsimonious. Moreover, many of the features would provide little predictive power, or are highly correlated with other features. Both low information features and correlated features not only expand the size of the model, but may also lead to biased estimates. Therefore, we take a two step approach for identifying relevant features. First, we remove low information features (described above), and second, we use a regularized regression strategy to identify the concise set of features, while still ensuring good model performance.

There are a number of different regularized regression techniques that could be employed (Hoerl 1970, Santos 1986, Zou 2005), with each introducing a penalty on the size of the coefficients in the model, inducing coefficients to be smaller and less likely to be statistically significant. We use the Elastic net model (Zou and Hastie, 2005), which introduces both a linear combination of $L_1$ and $L_2$ penalty terms on the coefficients in the regression. A tuning parameter λ controls the strength of the regularization and allows for a family of models of increasing parsimony to be fit. Zou and Hastie (2005) provide a method for fitting coefficients that is less likely to introduce bias into coefficient estimates, which we employ here.

Applying the first step of our feature selection process reduces the feature set considerably (see Table 1). The remaining features are then used to fit an Elastic net model with increasing values of λ. For each model estimated we calculate several performance metrics including the Precision/Recall Area Under the Curve (PR AUC) and the Bayesian Information Criteria (BIC). We wish to maximize the value of the PR AUC and minimize the value of the BIC. We use these metrics to identify the most parsimonious model that still provides sufficient prediction.

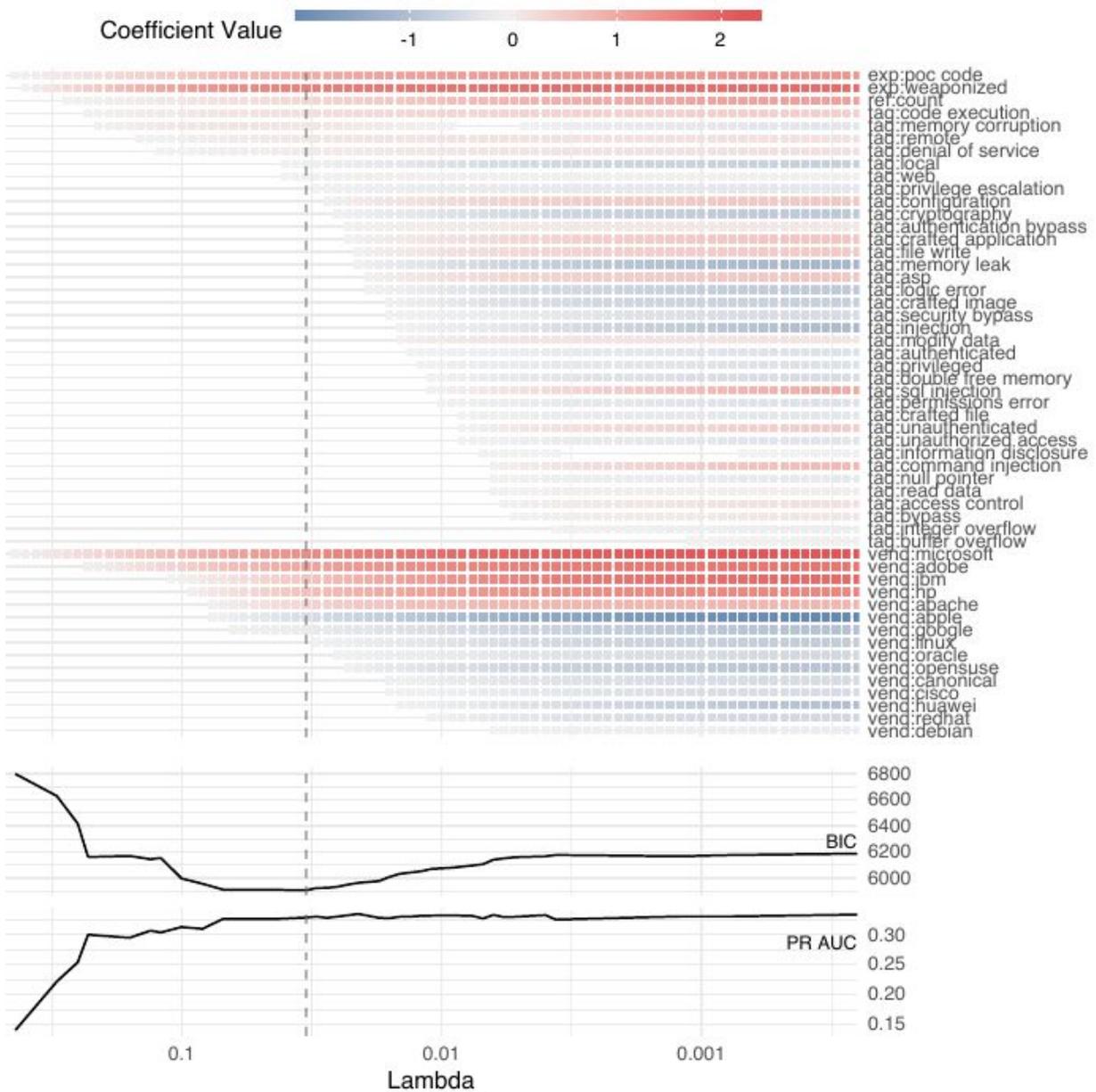

Figure 1: Results from elastic net regularization with BIC and PR AUC metrics

Note: Upper Panel: a red (blue) colored feature reflects a positive (negative) correlation with probability of exploit. The color intensity reflects the strength of the coefficient estimate (not statistical significance). Lower Panel: the BIC and AUC curves show their respective performance metrics along the y-axes. The x axis in all plots represents the number of variables considered.

The results of this process can be seen in Figure 1. The upper panel shows the 53 most significant features based on all available data and after filtering (2 exploit features, 36 tag features, and 15 vendor features). The vertical dashed line in Figure 1 shows the optimal (minimized) value of BIC, which occurs with the first sixteen variables. This represents the best tradeoff between performance and parsimony. Adding additional variables beyond that increases the BIC with nominal increases in the PR AUC.

Of the final variables selected, 7 related to the software vendor (*VENDOR*: Microsoft, Adobe, HP, Apache, IBM, Apple, and Google), 2 related to exploit code (*EXPLOIT*: whether exploit code had been published, and whether exploit code had already been weaponized), and 6 related to properties of the vulnerability and impact (*TAG*: memory corruption, code execution, local, remote, and web), and the final variable is a count of the references (*REF*) in a published CVE. The final set of 16 variables are shown in Table 2 along with descriptive statistics for each of the variables. Again, the overall percent of exploited vulnerabilities within our dataset is 3.7% (n = 25,159).

Table 2: Descriptive Statistics

| Variable | Percent Exploited | N |
|---|---|---|
| POC Weaponized | 37.1% | 283 |
| Vend: Microsoft | 21.2% | 1,333 |
| POC Published | 16.5% | 2,212 |
| Vend: Adobe | 15.9% | 747 |
| Memory Corruption | 11.7% | 1,570 |
| Vend: HP | 11.0% | 308 |
| Vend: Apache | 9.3% | 354 |
| Vend: IBM | 7.2% | 1,202 |
| Code Execution | 7.1% | 7,278 |
| Denial of Service | 5.6% | 8,547 |
| Remote | 5.4% | 10,681 |
| Web | 4.9% | 5,317 |
| Local | 3.5% | 2,686 |
| Vend: Apple | 0.8% | 877 |
| Vend: Google | 0.8% | 1,852 |
| Ref. Count | Avg Exploited: 5.3% Avg Not Exploited: 3.7% | 25,159 |

Our estimating model therefore becomes,

$$\Pr[\text{exploitation}]_i = f(\alpha_0 + \alpha_v VENDOR_i + \alpha_T EXPLOIT_i + \alpha_E TAG_i + \alpha_R REF_i + \varepsilon_i) \quad \text{Eq. 1}$$

Where, *VENDOR* is a vector of binary (dummy) variables related to the most frequently exploited software vendors, as described above. *EXPLOIT* is vector of binary variables related to the exploit code, *TAG* is a vector of variables related to characteristics of the vulnerability itself and *REF* is the log value of 1 plus the count of references in the published CVE. $\varepsilon$ is the random error term, assumed to be independent of the observed covariates. Identification of the variables of interest comes from the portion of vulnerabilities for which exploits have been observed in the wild.

## Results

The results of estimating Eq. 1 are shown in Table 3.

| Variable | LogOdds | Standard Error |
|---:|---:|---:|
| vend:microsoft | 2.44*** | 0.111 |
| vend:ibm | 2.07*** | 0.138 |
| exp:weaponized | 2.00*** | 0.164 |
| vend:adobe | 1.91*** | 0.136 |
| vend:hp | 1.62*** | 0.213 |
| exp:poc code | 1.50*** | 0.091 |
| vend:apache | 1.10*** | 0.231 |
| ref:count | 1.01*** | 0.078 |
| tag:code execution | 0.57*** | 0.096 |
| tag:remote | 0.23** | 0.089 |
| tag:denial of service | 0.22* | 0.098 |
| tag:web | 0.06 | 0.091 |
| tag:memory corruption | -0.20 | 0.126 |
| tag:local | -0.63*** | 0.143 |
| vend:google | -0.89** | 0.280 |
| vend:apple | -1.92*** | 0.399 |
| (Intercept) | -6.18 | 0.143 |

*Significance of p-value:  \*\*\* <0.001, \*\*<0.01, \*<0.05*

Table 3: Regression Results

These results are also displayed graphically in Fig 2.

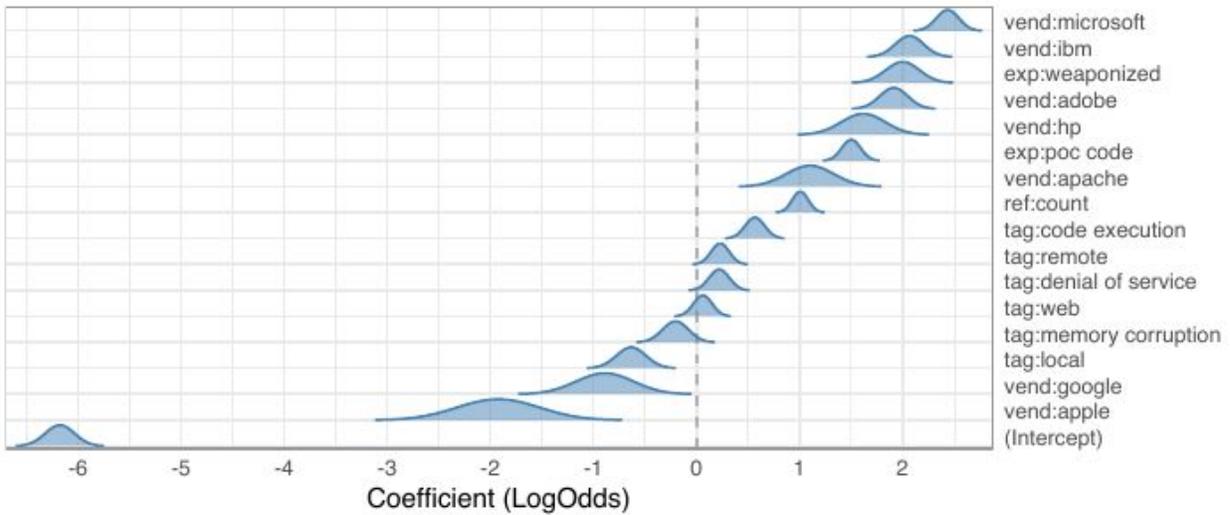

Fig 2: Graphical regression results

*Proof of concept* exploits being published and especially being *weaponized* (built into an exploitation framework) is clearly correlated to the probability that a vulnerability will be exploited. Among the vendors, the two most significantly correlated with the probability of exploitation are *Microsoft* and *IBM*. The *Microsoft* variable is likely a reflection of the ubiquitousness of Microsoft products (operating systems, and desktop and server applications), as well as a long history of being targeted for exploitation. The *IBM* variable appears to be related to a handful of exploited and widely used products being led by their flagship Websphere application. The *Adobe* variable appears due to its Acrobat product, which has had a long history of vulnerabilities. The *HP* variable is similar to *IBM* with a small handful of widely used products.

Vendor variables that were found to be negatively correlated with exploit include *Google* and *Apple*. The presence of *Google* is likely due to its adoption and integrated relationship with the CVE process which increases the amount of Google vulnerabilities that receive a CVE ID. The result of this integration is that Google has many published vulnerabilities (more than Microsoft) and yet very few are observed to be exploited. *Apple* is a closed platform and is traditionally less targeted by exploit writers. Vulnerability attributes (tags) tend to influence the probability of exploit less than other categories, but clearly a *remote code execution* increases the probability of exploit more than a *local memory corruption* vulnerability. Finally, the *count of references* published with a CVE is also positively correlated with the probability of exploit, and may have a strong effect. For example, CVE-2019-0708 (also known as "BlueKeep", which received a rare "cybersecurity advisory" from the NSA recommending the vulnerability be addressed) has 10 unique references published in the CVE. By comparison, just 4% of CVEs have 10 or more references, in our sample data set.

## Model Validation and Robustness Checking

The overall approach was validated by splitting the sample data by using a rolling forecasting origin technique (Hyndman and Athanasopoulos, 2013). This sets up a rolling window across the sample data where the training data always occurs prior to the hold-out test data. This is useful in the case of vulnerabilities because we want to validate how it performs on *future* vulnerabilities. One drawback is that we cannot use the full data set for validation as would be possible in a randomized cross-validation process (see below, we do this for comparison).

Figure 3 shows the performance of the model on the features discussed in the previous section with solid blue lines. The left panel in Figure 3 is the Receiver Operator Characteristic (ROC) curve which plots the false positive rate against the true positive rate and the more the curve pushes into the upper left, the better the performance of the model. We included a dashed line that represents a random strategy. On the right side of Figure 3 we are showing the Precision/Recall curve (efficiency/coverage). The more this line pushes to the upper right, the better the performance of the model. Again, there is a dashed line showing the performance of a purely random strategy.

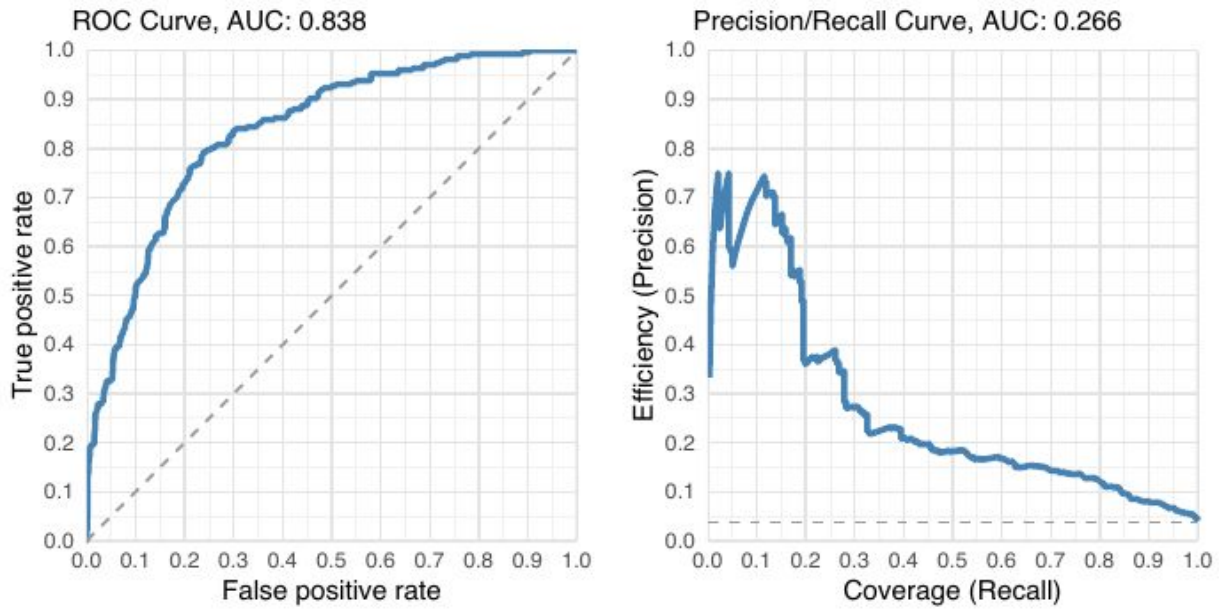

Figure 3: Model Performance with a time-based split

# Comparison to CVSS

For comparison, we can compute the true and false positives along with the true and false negatives for a purely CVSS strategy on our sample data set. From these metrics we can compute the true positive rate (TPR), false positive rate (FPR), efficiency and coverage. True positive rate is computed as the true positives over the amount of exploited vulnerabilities (true positive + false negatives) and the false positive rate is computed as the false positives over the non-exploited vulnerabilities (false positives + true negatives). Efficiency (precision) is computed from the true positives over the level of effort (true positives + false positives). Coverage is computed as the True Positives over the amount of exploited vulnerabilities (true positive + false negatives).

| Base Score | False Negatives | False Positives | True Negative | True Positive | TPR | FPR | Efficiency (Precision) | Coverage (Recall) |
|---|---|---|---|---|---|---|---|---|
| CVSS 10+ | 235 | 391 | 5,883 | 41 | 15.1% | 4.6% | 9.5% | 14.9% |
| CVSS 9+ | 207 | 744 | 5,530 | 69 | 31.5% | 12% | 8.5% | 25.0% |
| CVSS 8+ | 141 | 1,517 | 4,757 | 135 | 51% | 25.2% | 8.2% | 48.9% |
| CVSS 7+ | 103 | 2,600 | 3,674 | 173 | 64.9% | 43.8% | 6.2% | 62.7% |
| CVSS 6+ | 99 | 3,110 | 3,164 | 177 | 67.3% | 49.9% | 5.4% | 64.1% |
| CVSS 5+ | 58 | 4,248 | 2,026 | 218 | 79.3% | 67.2% | 4.9% | 79.0% |
| CVSS 4+ | 9 | 5,974 | 300 | 267 | 96.4% | 94.8% | 4.3% | 96.7% |

Table 4: Measurements for CVSS performance

This now sets up a direct comparison between EPSS and CVSS. Figure 4 repeats the ROC curve and PR curve from Figure 3, but includes the computed metrics (in red points) for several different CVSS-based strategies. Our model is clearly outperforming CVSS on both sets of metrics.

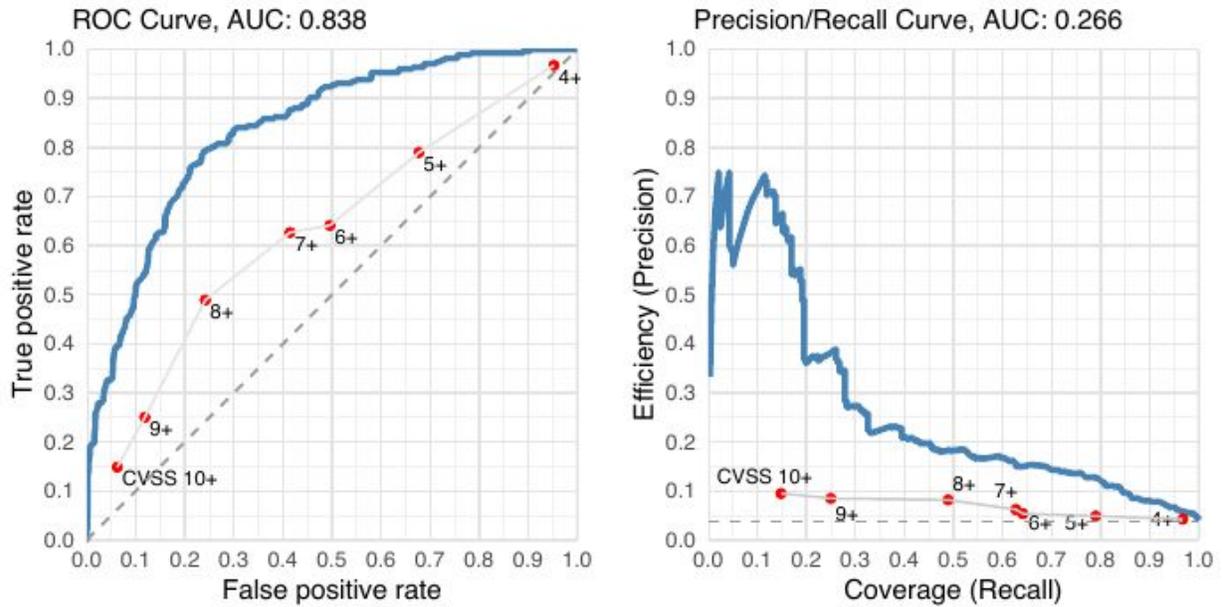

Figure 4: Model Performance with a time-based split and CVSS performance

Since most security practitioners (and companies) would like to maximize coverage in the most efficient way. We can compare the effort it would take to match the coverage of a CVSS-based strategy against the performance of EPSS. The level of effort in the CVSS approach can be found by adding the True Positive and the False Positives.

| Base Score | Coverage | CVSS Effort | EPSS Effort | EPSS Cutoff[11] | Reduction in Effort |
|---|---|---|---|---|---|
| CVSS 10+ | 14.9% | 432 | 61 | 44.1% | 85.9% |
| CVSS 9+ | 25.0% | 813 | 181 | 19.8% | 77.7% |
| CVSS 8+ | 48.9% | 1,652 | 738 | 5.8% | 55.3% |
| CVSS 7+ | 62.7% | 2,773 | 1,091 | 3.68% | 60.7% |
| CVSS 6+ | 64.1% | 3,287 | 1,174 | 3.35% | 64.3% |
| CVSS 5+ | 79.0% | 4,466 | 1,712 | 1.71% | 61.7% |
| CVSS 4+ | 96.7% | 6,241 | 4,443 | 0.54% | 28.8% |

Table 5: Level of effort for equivalent coverage between our model and CVSS

---

[11] The "cutoff" is the value at which all vulnerabilities above would be prioritized and all those below would be delayed.

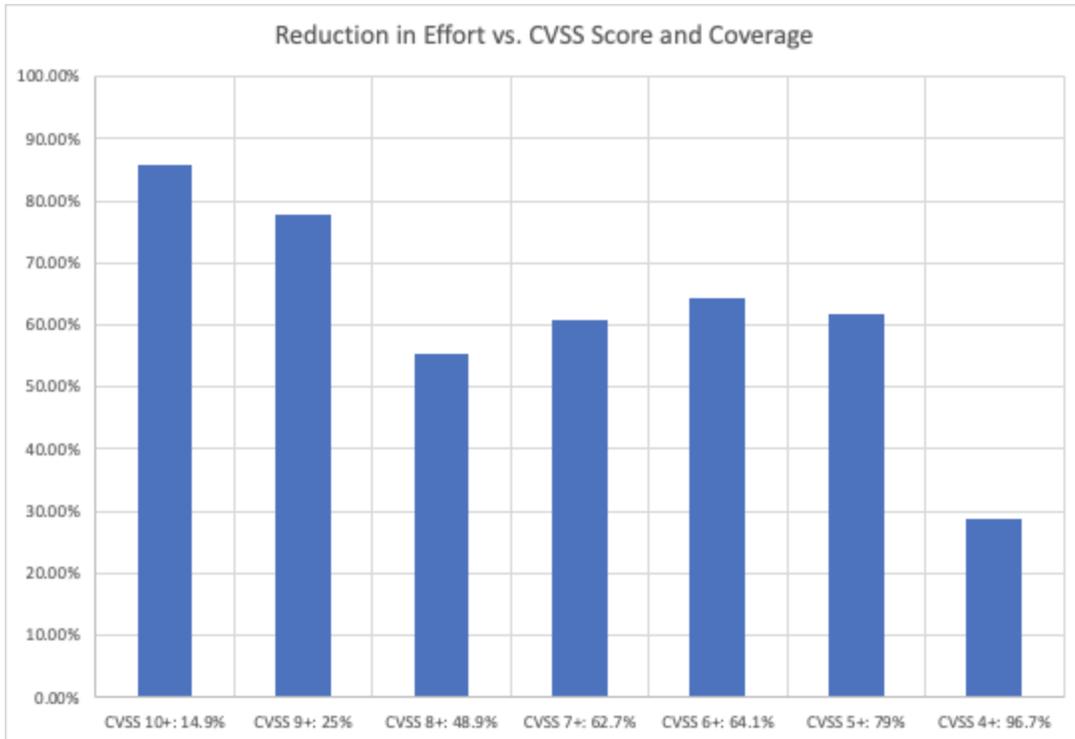

Figure 5: Reduction in Effort, matching Coverage: EPSS vs CVSS Score

## Calibration and Distribution

Given that the output is a probability, we can construct a calibration plot (sometimes referred to as a reliability plot). By binning the predicted probability and measuring the proportion of those observed to be exploited we generate the left-hand side of figure 6. The dashed line represents perfect calibration (when the model predicts 20% chance of exploitation we should expect 20% of those to be exploited). As figure 6 shows, the model is fairly well-calibrated. The predicted values (solid line) follow the calibrated line (dashed line), and generally fall within the 95% confidence interval in the lightly shaded region of the calibration plot.

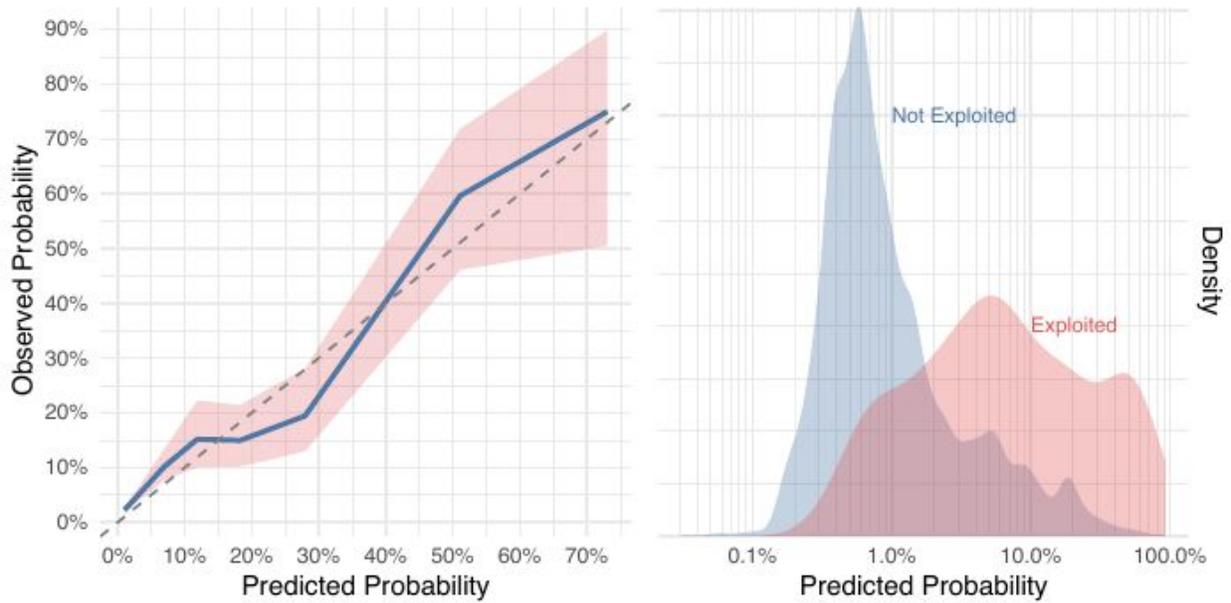

Figure 6: Reduction in Effort, matching Coverage: EPSS vs CVSS Score

The right hand side of Figure 6 shows the overall distribution of predicted values as a density plot. Note this is on the log scale to amplify the lower end of the prediction range as 76% of the predictions are 2% or less. While the density shows that there is a clear separation in the estimated probabilities between vulnerabilities that were observed to be exploited and those that were not, it is not a perfect separation. The description statistics about the distribution of predicted values is shown in Table 6.

|  | Min. | 1st Quartile | Median | Mean | 3rd Quartile | Max |
|---|---|---|---|---|---|---|
| Overall | 0.03% | 0.4% | 0.8% | 3.0% | 1.9% | 92.1% |
| Not Exploited | 0.03% | 0.4% | 0.7% | 2.5% | 1.6% | 92.1% |
| Exploited | 0.3% | 2.1% | 5.7% | 15.4% | 19.8% | 80.7% |

Table 6: Summary statistics for the distributions of predicted values

## Randomized 5-fold Cross-Validation

While randomized cross-validation can generate predictions across the entire sample data, it may be misleading with time-series data as we discussed previously. If the underlying system is shifting over time, historical observations may not be strong predictors of future outcome. We include this here as a comparison and to explore the outcome if time were not factored into model validation. Figure 7 shows a five-fold cross-validation for both the ROC and AUC curve. Again, the CVSS-based approach is shown as points on the same scale.

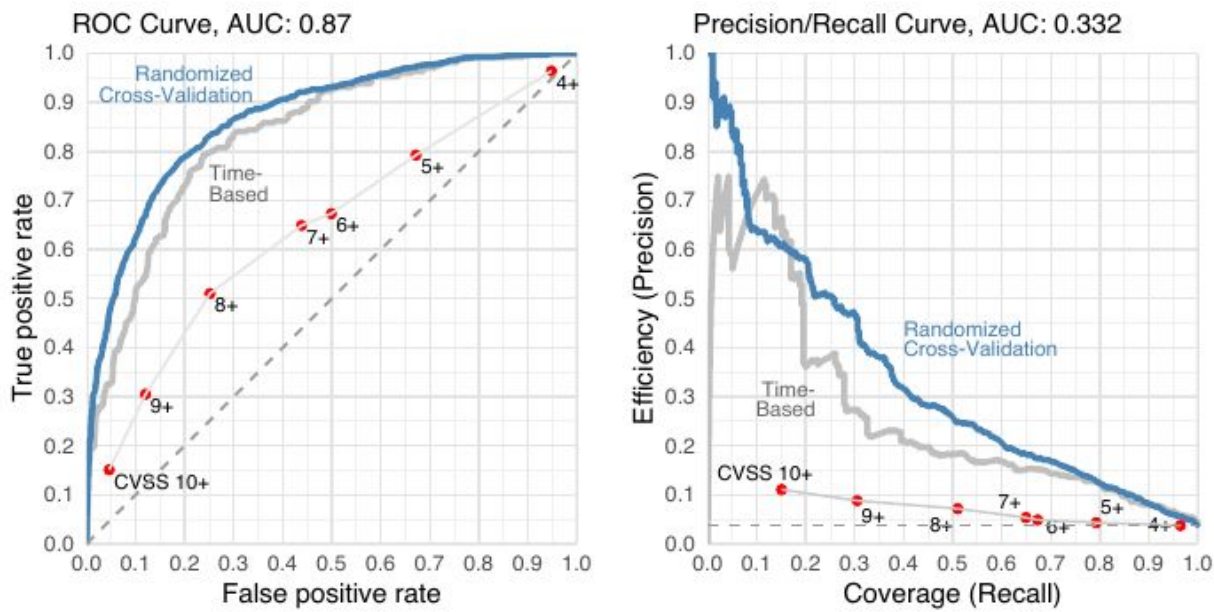

Figure 7: Model Performance with randomized cross-validation splits and CVSS performance

## Focusing on Existing Vulnerabilities

Because Kenna Security works with hundreds of enterprises - in their vulnerability management programs, we can see that not all CVEs are a concern for companies. In fact, only 44% of the CVE's published in the two-year time window of this study were ever reported as open in a corporate environment. This occurs for two reasons: not all software is used by enterprises, and not all vulnerabilities have scanner signatures written for them. The subset we study here are CVEs which have detection signatures written by vulnerability management scan vendors, and have occurrences of these scans in the wild. By focusing on just the vulnerabilities that the vulnerability scanners are reporting to companies we may get a more realistic picture of what companies could experience by implementing EVSS.

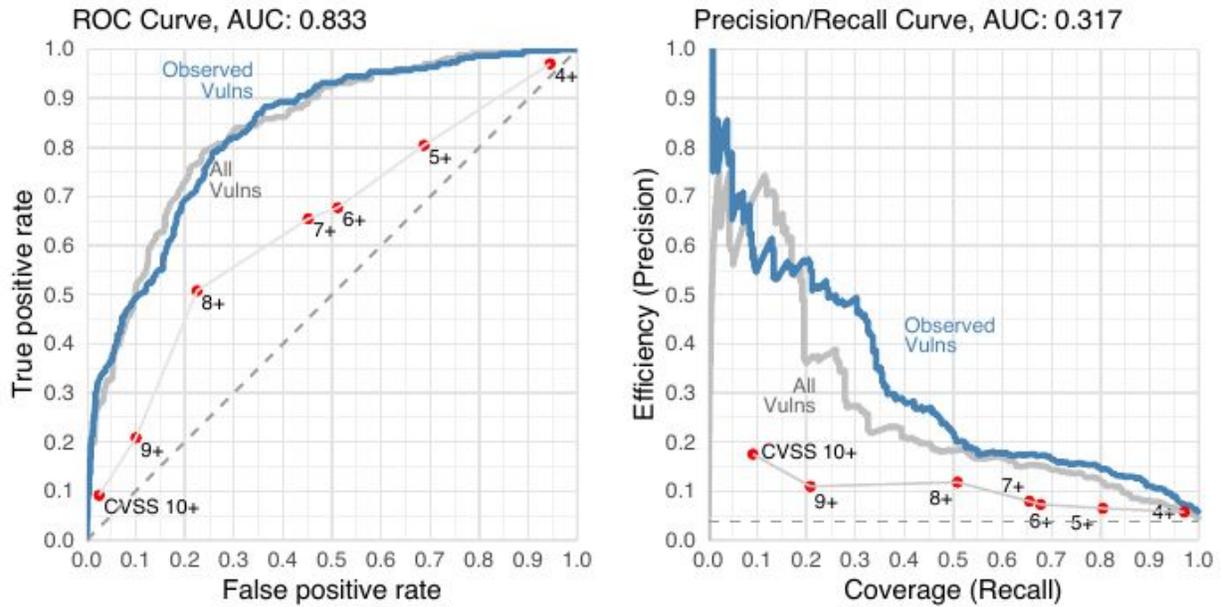

Figure 8: Model Performance with time-based splits on open/scanned vulnerabilities

The ROC curve on the left does not see much of an improvement, but the PR curve on the right does get some lift in the middle portion of the curve. One reason for this could be that the scanner vendors and corporate environments are focused on popular vendors and technologies, for which we have robust data, and more data earlier on.

# Formalizing and Implementing the EPSS

Next we consider an example vulnerability and we compute the estimated probability of exploitation within the first 12 months after publishing. First step is to compute the log odds which is a cumulative sum of the observations about a vulnerability multiplied by the coefficients from the model. Therefore, using the results presented in Table 3, the log odds can be computed as,

$$
\begin{aligned}
\text{LogOdds} = -6.18 + \\
2.44 \ * \ \text{vend:microsoft} + \\
2.07 \ * \ \text{vend:ibm} + \\
2.00 \ * \ \text{exp:weaponized} + \\
1.91 \ * \ \text{vend:adobe} + \\
1.62 \ * \ \text{vend:hp} + \\
1.50 \ * \ \text{exp:poc code} + \\
1.10 \ * \ \text{vend:apache} + \\
1.01 \ * \ \log(\text{ref:count} + 1) + \\
0.57 \ * \ \text{tag:code execution} + \\
0.23 \ * \ \text{tag:remote} + \\
0.22 \ * \ \text{tag:denial of service} + \\
0.06 \ * \ \text{tag:web} + \\
-0.20 \ * \ \text{tag:memory corruption} + \\
-0.63 \ * \ \text{tag:local} + \\
-0.89 \ * \ \text{vend:google} + \\
-1.92 \ * \ \text{vend:apple} \quad\quad\quad \text{Eq. 2}
\end{aligned}
$$

Where each variable on the right hand side of Equation 2 is encoded as a 1 or 0 depending on if the attribute is present (1) or not (0) in the vulnerability, with the exception of the reference count which is a continuous variable transformed by adding 1 and taking the log (adding one accounts for the possibility that the reference count may be zero).

Additionally, the LogOdds value is converted into the estimated probability of exploitation as,

$$\Pr[\text{exploitation}] = 1/(1+e^{-\text{LogOdds}}) \quad\quad\quad \text{Eq. 3}$$

For example, consider CVE-2019-0708, also known as "BlueKeep", which was published Nov 26th, 2018 (several months after our sample window). It was (is) a widely discussed weakness in Microsoft's Remote Desktop that allow remote code execution with attributes as shown in Table 7.

Table 7: Regression Results for CVE-2019-0708

| Variable | Coefficient | Observations | |
|---|---|---|---|
| (Intercept) | -6.18 | 1 (always) | -6.18 |
| vend:microsoft | 2.44 | 1 | 2.44 |
| vend:ibm | 2.07 | 0 | 0 |
| exp:weaponized | 2.00 | 1 | 2.00 |
| vend:adobe | 1.91 | 0 | 0 |
| vend:hp | 1.62 | 0 | 0 |
| exp:poc code | 1.50 | 1 | 1.50 |
| vend:apache | 1.10 | 0 | 0 |
| ref:count | 1.01 | log(10 + 1) = 2.4 | 2.424 |
| tag:code execution | 0.57 | 1 | 0.57 |
| tag:remote | 0.23 | 1 | 0.23 |
| tag:denial of service | 0.22 | 0 | 0 |
| tag:web | 0.06 | 0 | 0 |
| tag:memory corruption | -0.20 | 0 | 0 |
| tag:local | -0.63 | 0 | 0 |
| vend:google | -0.89 | 0 | 0 |
| vend:apple | -1.92 | 0 | 0 |
| | | sum: | 2.984 |

Simplifying Table 5 into a singular equation,

$$\text{log odds} = -6.18 + 2.44 + 2 + 1.5 + 2.424 + 0.57 + 0.23 = 2.984.$$

Now we apply Equation 3,

$$\Pr[\text{exploitation}] = 1/(1+e^{-2.984}) = 1 / (1 + 0.05059007) = 0.952 = 95.2\%$$

Therefore this model predicts that the probability of CVE-2019-0708 being exploited within 12 months of being published is approximately 95%. As a point of reference, this CVE was the single highest rated in the twelve months following the data sample used in model training.

# Discussion and Limitations

We are striving to create an approach that is 1) simple to implement 2) implementable with open data 3) interpretable 4) parsimonious and 5) performant. We necessarily had to make concessions based on limitations in the data and the practicalities of implementation by keeping the variables included as small as possible.

Necessarily, the model we present here is built using outcome data (exploitation in the wild) that is not freely available. The predictions are based on observed exploit data identified by standard signature-based intrusion detection systems, and may therefore be subject to missed exploit activity. That is, we do not observe, and therefore cannot make predictions about, exploits that were launched, but not observed. There are any number of reasons why an exploit may not be observed: no IDS signature had not been created or it occurred in locations or networks where we had no visibility. We are also limited to the time window of vulnerabilities used. Given that we bound the prediction to a 12 month window, we necessarily must restrict data collection to omit observations newer than this period.

We use a variety of sources to get as close as possible to the most holistic picture of exploitation. The closed (and closely guarded nature) of exploitation data prevents us from using fully open data to measure exploitation. Our goal is to create an implementable, model which accepts open and freely available inputs and not necessarily one that could be trained from scratch using open data. If security practitioners, firms, or security vendors work with vulnerabilities, we encourage them to contribute to this or similar research efforts.  Improvements in data collection, specifically around signature generation, deployment and activity will have a direct and correlated improvement on the accuracy of the models.  Additionally, improved data collection about the vulnerabilities and the context in which they exist should also improve future modeling efforts. Lastly, additional exploitation event data should be correlated to CVE, allowing researchers to combine outcome events into larger datasets.

The EPSS model is attempting to capture and predict an outcome within an ever-evolving and complex system.  Despite this complex and ever evolving landscape, EPSS makes remarkably good predictions with a simple, interpretable linear model. Its simplicity means that it is easy to distribute and be built into existing infrastructures. EPSS's structure also allows for the easy creation and interpretation of counterfactuals, e.g. "how much do we expect the probability to increase if proof of concept code is released?". It is clear that any model of vulnerability exploitations will need to evolve with the underlying vulnerability landscape. EPSS will likely experience a decay in performance as time passes and will require updates and retraining over time. But with time and collaboration, there exists opportunities for the generation of specific models.  Research focused around specific vendors or on classes of vulnerabilities would yield a more accurate prioritization system.

We believe a key advantage to EPSS is that it can be augmented and improved. EPSS is created to be as parsimonious as possible while still maintaining performance, but that doesn't preclude future augmentation. Augmentation can come in the form of more sources of exploitation, new vulnerability features, and longer time periods of analysis. As more kinds of disparate data – data for which we have not even considered – become available, this could allow us to further refine and improve the model, identifying either new or stronger correlations. For example, it is conceivable that data from social media platforms, or vulnerability scan data from additional corporate networks could provide useful inferences. future research in machine learning techniques may provide model techniques and allow us to make better coefficient estimates.

We are clear to restrict the context of our scoring system to provide estimates of threats, rather than true risks. That is, we recognize that while the severity of a vulnerability is characterized by CVSS, and this scoring system characterizes the probability that a vulnerability will be exploited, neither of these, nor the combination of them, represent a complete measure of risk because we do not observe nor incorporate firm-level information regarding a firm's assets, its operating environment, or any compensating security controls. In addition, we provide no information regarding the cost of patching vulnerabilities, as these are firm-dependent expenses.

We only consider vulnerabilities that have been assigned CVE identifiers since the CVE identifiers represent a common identification mechanism employed across our disparate data sources. Adoption and/or consideration of alternative vulnerability databases was not feasible as the data collection sources did not include references to them. While we would have liked to account for and model all vulnerabilities discovered, the lack of a common identification method made data aggregation improbable. As a result, we omit other kinds of software (or hardware) flaws or misconfigurations that may also be exploited. Though, it is possible that these may be incorporated into future versions of this scoring system if data sources enabled it.

It is conceivable that by disclosing information about which vulnerabilities are more likely to be exploited, based on past information, that this may change the strategic behavior of malicious hackers to select vulnerabilities that would be less likely to be noticed and detected, thereby artificially altering the vulnerability exploit ecosystem. We believe this is suggestive, at best, but is something that we will seek to identify and minimize.

# Conclusion

This work undoubtedly represents an early step in vulnerability prioritization research. Hopefully the benefit of this work has become evident and may motivate additional research in this area. Vulnerability research will (and should) begin with the data. EPSS was conceived out of a recognition by experienced security practitioners and researchers that current methods for assessing the risk of a software vulnerability are based on limited information that is, for the most part, uninformed by real-world empirical data. For example, while CVSS is capturing the immutable characteristics of a vulnerability in order to communicate a measure of *severity*, it has been misunderstood and misapplied as a measure of *risk*. In some sense, however, this could have been expected. Humans suffer from many cognitive biases, which cause us to apply basic heuristics in order to manage complex decisions.[12] And this craving for a simplistic representation of information security risk is often satiated when we're presented with a *single number,* ranging from 0 to 10.

But we must do better. We must understand that security risk is not reducible to a single value (a CVSS score). Nor is the entirety of security risk contained in both CVSS and EPSS scores. But so far this is what we have. Until we acquire new data, techniques and/or methods, we must consider how EPSS (the probability of exploitation) and CVSS (an ordinal numerical scale that reflects a set of characteristics of vulnerability severity) may coexist. To that end, let us consider three approaches.

A first approach could be to substitute EPSS for CVSS entirely within all decision-making policies. Since they both produce bounded, numerical scores that could be scaled identically, substitution could be simple and straightforward. The concern with this approach, however, is that (as previously discussed) threat is also just one component of risk, and this strategy would ignore all the qualities captured by a severity score.

Second, a keen practitioner may be drawn to simply multiple the CVSS score by the EPSS probability in order to produce a number that measures the *severity * threat* of a vulnerability -- and hope this becomes a better (closer) measure of risk. While intuitively, this may feel appropriate, logically and mathematically, it is not. Applying mathematical operations to combine the probability of exploitation against an ordinal value is faulty and should be avoided.

A third approach would be to recognize that CVSS and EPSS each communicate orthogonal pieces of information about a vulnerability, and instead of mashing them together mathematically, to consider the values together, but separately. At a first approximation, it is clear that high severity *and* high probability vulnerabilities should be prioritized first within an organization. Similarly, low severity *and* low probability vulnerabilities could be deprioritized within an organization. What remains are the unclassified vulnerabilities that require additional consideration of the environment, systems and information involved. While likely unsatisfying as a complete risk management decision framework, we believe this is unavoidable, yet realistic.

---

[12] For example, consider bounded rationality (Simon, 1957), the notion that we find many ways of simplifying complex decision by making simplifying assumptions. And see https://www.visualcapitalist.com/every-single-cognitive-bias/ for a list of all documented cognitive biases. Last accessed July 25, 2019.

We, the authors of this paper, do not yet have the perfect or final solution. We recognize that enterprises are massive vessels, that have developed policies and practices to suit their culture and capabilities over many years. And changing course would take much effort, even when a preferred direction has been identified. However, augmenting existing policies with the information that EPSS provides, to any of the previously described degrees, can increase the efficiency of existing policies, and pave the way for systemic re-evaluations of these policies.

While we believe that EPSS provides a fundamental and material contribution to enterprise security risk management and national cyber security policymaking, the implementation of this threat scoring system will be an evolving practice. It is our genuine hope and desire, therefore, that EPSS will provide useful and defensible threat information that has never before been available.